\documentclass[10pt,aps,prx,showpacs,superscriptaddress,preprint,citeautoscript]{revtex4-1}

\newcommand{\de}{\delta}

\newcommand{\been}{\begin{equation}}
\newcommand{\een}{\end{equation}}
\newcommand{\beena}{\begin{eqnarray}}
\newcommand{\eena}{\end{eqnarray}}

\usepackage{amsfonts,amssymb,amsmath,amsthm,geometry}
\usepackage{graphicx}
\usepackage{psfrag}
\usepackage{color,xcolor}
\usepackage{subfigure}
\usepackage{float}

\newtheorem{theorem}{Theorem}
\newtheorem{lemma}[theorem]{Lemma}

\def\eeq{\relax}
\def\beq#1#2\eeq{\begin{equation}\label{#1}#2\end{equation}}
\def\bal#1#2\eal{\begin{align}\label{#1}#2\end{align}}
\def\bse#1#2\ese{\begin{subequations}\label{#1}#2\end{subequations}}
\def\ba{\begin{aligned}}   \def\ea{\end{aligned}}

\def\XXint#1#2#3{{\setbox0=\hbox{$#1{#2#3}{\int}$}
\vcenter{\hbox{$#2#3$}}\kern-.5\wd0}}

\def\dd{\operatorname{d}}
\def\de{\operatorname{e}}

\def\diag{\operatorname{diag}}
\def\i{\operatorname{i}}
\newtheorem{thm}{Theorem}

\def\dd{\operatorname{d}} 

\begin{document}
\def\singlespacing{\baselineskip=13pt}	\def\doublespacing{\baselineskip=18pt}
\singlespacing

\title{Static elastic cloaking, low frequency elastic wave transparency and neutral inclusions}

\author{Andrew N. Norris}
\affiliation{Mechanical and Aerospace Engineering, Rutgers University, Piscataway, NJ 08854-8058, USA}
\email{norris@rutgers.edu}
\author{William J.~Parnell}
\affiliation{School of Mathematics, University of Manchester, Oxford Road, Manchester, M13 9PL, UK}



\begin{abstract}
{New} connections between static elastic cloaking, low frequency elastic wave scattering and neutral inclusions are established in the context of two dimensional elasticity.  A cylindrical core   surrounded by a cylindrical shell  is embedded in a uniform elastic matrix.  Given the core and matrix properties, we answer the questions  of how to  select the shell material such that (i) it acts as a static elastic cloak, and (ii) it  eliminates low frequency scattering of incident elastic waves.  It is shown that static cloaking (i) requires an anisotropic shell, whereas  scattering reduction (ii) can be satisfied more simply with isotropic materials. Implicit solutions for the shell material are obtained by considering the core-shell composite cylinder as a neutral elastic inclusion. Two types of neutral inclusion are distinguished, \textit{weak} and \textit{strong} with the former equivalent to low frequency transparency {and the classical Christensen and Lo generalised self-consistent result for in-plane shear from 1979. Our introduction of the \textit{strong neutral inclusion} is an important extension of this result in that we show that standard anisotropic shells can act as perfect static cloaks, contrasting previous work that has employed ``unphysical'' materials.} 
The relationships between low frequency transparency, static cloaking and neutral inclusions provide the material designer with options for achieving elastic cloaking in the quasi-static limit. 
\end{abstract}
\maketitle


\section{Introduction}
{The ability to cloak a region of space so that an incident field or applied loading does not see or feel the presence of an object is of great interest in science and}
{engineering. Over the last two decades significant progress has been made in this field in the domains of electromagnetism \cite{Schurig06, Cai08}, acoustics \cite{cummer07, norris08} and flexural waves on thin plates \cite{Climente16,Zareei2017}}. Cloaking of elastic waves {however, even in the quasi-static limit} requires materials with properties that are, at present, unachievable.  According to transformation elasticity \cite{Brun09, Norris11a}  one needs solids that display a significant amount of anisotropy combined with strong asymmetry of the elastic stress.  Large anisotropy is common in composite materials and can be engineered by design, but significant stress asymmetry is not seen in practical materials. Some mechanisms to circumvent this difficulty have been proposed, including isotropic polar solids for conformal transformation elasticity and cloaking \cite{Nassar2019,Nassar2019b},  hyperelastic materials under pre-stress \cite{Parnell2011, Norris2012aa}, and under some  circumstances solutions can be found in the case of thin plates that do not need asymmetric stress \cite{Col-14, Liu-16}. {Recent work has employed lattice transformations to cloak in-plane shear waves \cite{kadic20}.}


Restricting attention to statics on the other hand, a purely static cloak is an elastic layer that has the effect of ensuring that the deformation exterior to the cloaked region is the same as if there were no object or layer present.    
Static cloaking is closely related with the concept of a {\it neutral inclusion} (NI), which is a region of inhomogeneity in an otherwise uniform solid that does not disturb an applied exterior field. {NIs can be tailored to specific loading types, whereas a static cloak will ensure that there is no influence to the presence of an object for \textit{any} type of imposed field.} NIs are therefore {by definition statically cloaked for a certain imposed field}.  Examples of NIs are Hashin's 
coated sphere \cite{Hashin1962}   for   conductivity, later generalised to coated confocal ellipsoids   \cite{Milton1981}  and other possible
shapes  \cite{Milton01a}.  
  The associated scalar potential problem and associated NIs and coated NIs have been studied extensively, see \cite[\S7]{Milton01}, \cite{Benveniste2003} and references therein.
Extensions to the case of nonlinear conductivity \cite{jimenez2013nonlinear}  and hydrostatic loading in plane finite elasticity have also been considered  \cite{Wang2012}.
 The two-dimensional scalar potential problem is  pertinent in the   context of the anti-plane elastic problem \cite{Milton01a, Kang14}.  The full elastostatic NI problem is more challenging and there have been a number of relevant studies in linear elasticity \cite{Bigoni1998,Ru1998, He2002,Benveniste2003, Bertoldi2007,Wang2012a, Kang16}. A general elastic NI, or a condition to realise one, does not appear to have been exhibited with finite thickness shells, {although see \cite{Benveniste2003} where NIs are derived for special loading scenarios}.  Instead it is often the case that ``interface'' type conditions are required for the combined shear/bulk modulus neutrality, i.e.\ for neutrality to be achieved under general in-plane loadings \cite{Ru1998}. 
 There has been some success in realization of an approximate core-shell design based on Hashin's assemblage using a pentamode material for the shell, a so-called unfeelability cloak \cite{Buckmann14b}.

\begin{figure}[h]
\begin{center}
	\includegraphics[width=0.5\linewidth]{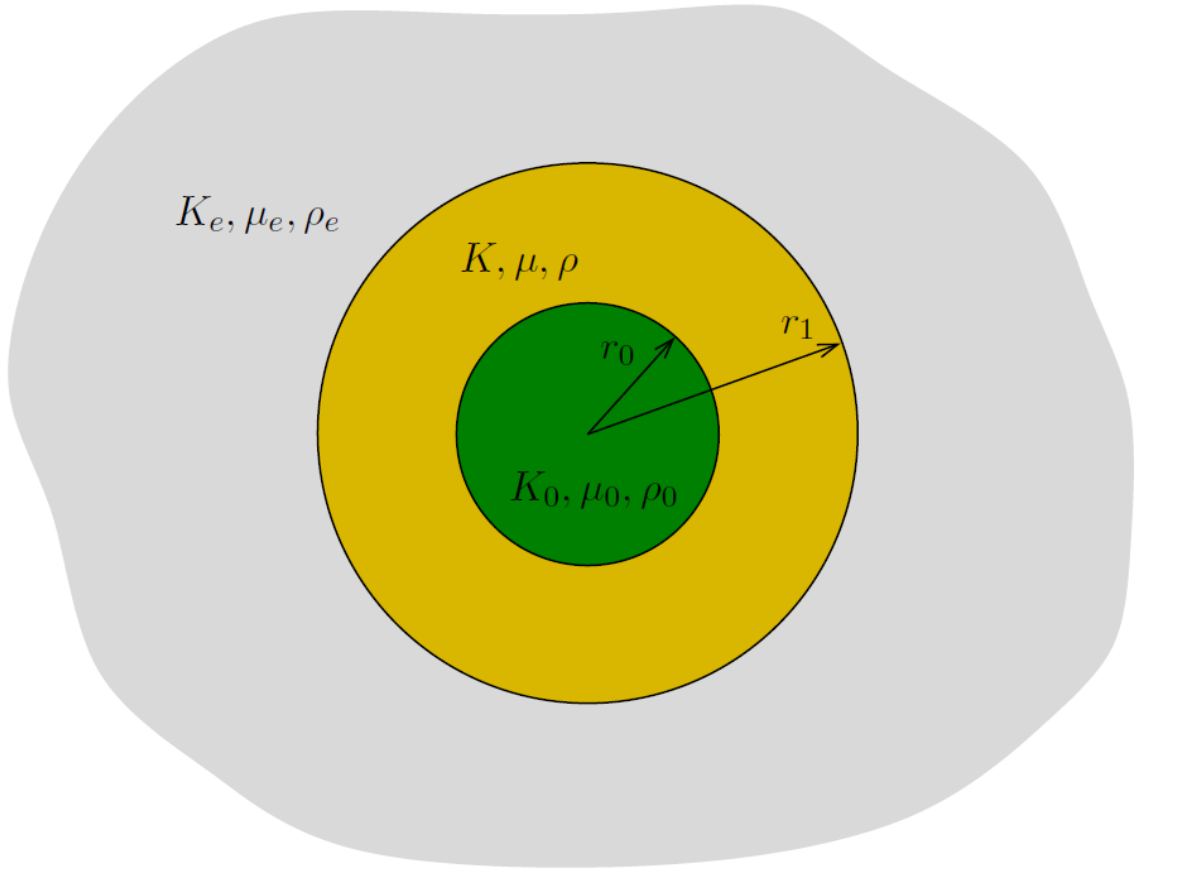}
\end{center}
\caption{The central cylindrical core of radius $r_0$ is surrounded by  a cylindrical shell (or layer or annulus) of elastic material of outer radius $r_1$.  The core density, shear modulus and in-plane bulk modulus are $\rho_0$, $\mu_0$ and $K_0$, the shell properties can be anisotropic and radially dependent, {although only the isotropic shell case is illustrated here}.  The composite cylinder lies in  an infinite uniform isotropic elastic medium, $\rho_e$, $\mu_e$ and $K_e$. }
\label{fig1}
\end{figure}

Our interest here is  two-dimensional {cylindrical} and {inhomogeneous NIs} for elasticity.  {As motivation, consider a cylindrical core region 
 surrounded by a shell (or coating, layer, annulus)
all of which is embedded in a host exterior medium, as illustrated in Figure \ref{fig1}.} 
The properties of the shell (homogeneous or inhomogeneous) are chosen so that the combined core and coating act as a NI. Unlike the cloak of transformation elastodynamics \cite{Norris11a}, the moduli of the static cloak will depend upon the properties of the cloaked object. 
The three-phase {elastically isotropic} configuration depicted in Figure \ref{fig1} has been studied previously, under the action of various far-field loadings in the context of estimating the static effective properties of core unidirectional fibres dispersed in a matrix with the properties of the coating.  By inserting the composite cylinder (core plus coating) in the background medium 
and requiring that it act as a NI, the background properties provide a \textit{self-consistent} estimate of the effective material properties, i.e.\ the matrix properties.  This approach partitions into two sub-problems, first for in-plane hydrostatic loading which ensures a condition can be determined for the effective bulk modulus. However the in-plane shear problem is under-determined; the ``perturbed'' static displacement \textit{outside} the composite cylinder depends on two coefficients, while the background material has only a single parameter: the effective shear modulus.  It is not possible to make both perturbed displacement coefficients vanish simultaneously, i.e.\ the composite cylinder cannot be a static cloak \textit{if the coating is isotropic}. 
As an alternative, Christensen and Lo  \cite{Christensen79} assumed that the strain energy
in the composite cylinder must be the same as the strain energy in an equivalent volume of the effective material, which can be satisfied by setting only one of the displacement coefficients to zero {(the stress terms associated with $1/r^4$ decay do not contribute to the strain energy)}. This procedure has since been termed the {Generalised Self Consistent Scheme} (GSCS) \cite{Hashin1990}.   The GSCS energy equivalence method has since been generalised to the case of multiply layered cylinders using transfer matrices \cite{Herve95}. The effect of anisotropy in fibres and coatings was considered by \cite{Avery1986} in relation to thermal properties of composites. Thermoelastic effective properties for orthotropic phases were derived using a combination of the GSCS and Composite Cylinders Assemblage (CCA) methods in \cite{Hashin1990}. A Mori-Tanaka inspired interaction approach was used by \cite{Chen1990} to consider thermomechanical loading of cylindrically orthotropic fibers with transversely isotropic coatings. The solutions were subsequently applied to  estimate effective properties of coated cylindrically orthotropic fiber reinforced composites \cite{Benveniste1991}. Recent reviews of relevance include  \cite{Charalambakis2010} on homogenization and micromechanics and \cite{Zhou2013, Par-16} on inclusions.

Christensen and Lo's solution for the composite cylinders model \cite{Christensen79} and its generalizations \cite{Hashin1990,Herve95} can be considered as {\it weak neutral inclusions} because the perturbed exterior field is not completely eliminated {(the $1/r^4$ decay in the exterior stress remains)} as compared with {\it strong neutral inclusions} for  which the exterior displacement and stress are unperturbed. A related but apparently quite distinct situation arises with scattering of time-harmonic elastic waves. The scattered, i.e.\ perturbed, exterior field, can be expressed as an asymptotic  series in a non-dimensional parameter proportional to the frequency.   We say that the scattering object is {\it transparent at low frequency} if both the leading order longitudinal and transverse scattered waves vanish.  The lowest order terms in the power series vanish for both the scattered longitudinal and transverse waves if the scatterer is a  neutral inclusion.  However, as discussed above,  a given two phase composite cylinder with isotropic phases can at most be a weak NI, 
which begs the question of how the weak NI relates to low frequency transparency.  

{One lesson taken from the composite cylinders model \cite{Christensen79} is that the shell must be anisotropic if strong NIs are to be obtained.  One of our main results is therefore to identify the type of anisotropy necessary to achieve static cloaking, or equivalently a strong NI effect. 
The static elastic cloak determined here is however distinct from others in elasticity \cite{Brun09, Norris11a,Parnell2011, Norris2012aa,Nassar2019,Nassar2019b} in that it can be realised using ``normal'' elastic materials corresponding to symmetric stress.}

\subsection{Objective and overview}  

The problem considered is as follows: given a host matrix and a cylindrical core, determine the shell properties such that the {core+shell (composite assemblage)} acts as either a strong or a weak neutral inclusion (NI), {to planar deformation of hydrostatic and shear loading}. Figure \ref{fig1}.  {We define the following terms.}
\begin{itemize}
\item \textit{Strong neutral inclusion}: the perturbed field exterior to the NI is 
zero.  This is equivalent to  a static elastic cloak.   
\item \textit{Weak neutral inclusion}: the strain energy of the NI is the same as the strain energy of an equivalent volume of matrix material. 
\end{itemize}
The core and matrix have material properties $K_0$, $\mu_0$, $\rho_0$ and 
$K_e$, $\mu_e$, $\rho_e$, respectively, {the \textit{in-plane} bulk modulus, shear modulus and density}.  Given the core radius $r_0$ and the outer radius of the shell $r_1$, the objective is to find properties of the shell that result in a NI of either type.  At the same time we are interested in the relation between NI effects, weak and strong, and low frequency transparency. 

We will explicitly show that a two-phase composite cylinder with  isotropic core and shell   cannot be a strong NI. An  isotropic shell can only yield a weak NI, and equations for the required properties $K$, $\mu$, $\rho$ will be obtained.  A strong NI requires that the shell be anisotropic, and the requisite conditions will be found.  
It will be shown that the NI and transparency  properties are related: a weak NI is transparent at low frequency, that is, both of the leading order scattered waves (longitudinal and transverse) vanish.  Conversely, low frequency transparency implies that the scatterer acts as a NI, weak or strong, but generally weak.

Given a composite cylinder and its properties one can ask what are the properties of the matrix which makes it act as a NI.  This is a standard \textit{effective medium} problem, which may be solved in an approximate or exact manner, as we will see in Sections \ref{sec2} and \ref{sec5}, respectively. 
Finding solutions for the NI layer properties is therefore an inverse problem: we will first solve the effective medium problem, with the NI properties {determined as} implicit functions of the core and matrix properties.  The approximate effective medium solution in Section \ref{sec2} provides the only explicit examples for the NI properties.  

Our approach to the exact solution of the NI problem combines static and dynamic solutions in a novel manner.     Unlike previous derivations of the effective bulk modulus, which require full solutions for the displacement and stress fields in the composite cylinder \cite{Christensen79, Herve95},  here it is found directly as the solution of an ordinary differential equation (ODE) of Riccati type.  The effective shear modulus involves a 2$\times$2 impedance matrix which satisfies a Riccati ODE. This matrix yields both the low frequency transparency and the NI conditions.  The former is derived using a low frequency expansion of the scattered field, giving a condition identical to the GSCS.  The NI condition for shear is a purely static one which reduces to a single constraint on the elements of the impedance matrix. In particular, we derive a simple condition which is necessary and sufficient to obtain a strong NI.
We provide examples of composite cylinders comprising isotropic cores and uniform anisotropic shells that are strong NIs {and illustrate two of these cases graphically showing the difference between the weak and strong NI in the process}.

The outline of the paper is as follows.  An approximate effective medium solution is used in Section \ref{sec2} to solve for (approximate) NI parameters.  The explicit solution shows that the range of possibilities decreases to zero in certain parameter regimes. Section \ref{sec3} outlines the exact forward solution approach for the composite cylinder effective medium problem, and relates the NI effect to low frequency transparency effects.   By representing the fields in terms of angular harmonics, it is apparent that there are two distinct problems to solve: for $n=0$ and $n=2$.  Solutions of the effective medium problem are given in Section \ref{sec4} for the effective bulk modulus ($n=0)$, and in  Section \ref{sec5}  for the effective shear modulus $(n=2)$.  Distinction between weak NI, strong NI and low frequency transparency become apparent in Section \ref{sec5}, where the exact NI solution is described.  {Examples of strong NI core-shell composite cylinders are presented in Section 6.}  Concluding remarks are given in Section \ref{sec7}. 

\section{Quasistatic cloak using an approximate model}  \label{sec2}

As an introduction to the problem we first demonstrate how one can use an \textit{approximate} model to estimate the properties necessary for an approximate static cloak. {It should be stressed that since the model is approximate the configuration cannot be classified as an exact neutral inclusion of any type, weak or strong. However it gives an indication of what is required of such a NI and its possible regimes of validity.} Consider the single core configuration as depicted in Fig.\ \ref{fig1}. In the context of effective medium theory the core (subscript $0$) and coating properties, together with core volume fraction $f\in (0,1)$, are given and the {external} properties (subscript $e$) are then determined subject to some consistency constraint. The static cloak problem is different in that the core and surrounding medium properties are given and the cloak (coating) properties are chosen in order to render either equivalent energy or zero transparency, etc.

%
As an example, let us use effective property estimates based on a modification of the Kuster-Toks\"oz model,   \cite[eq.\ (4)]{Wu2007}
\bse{-4}
\bal{4a}
\rho - \rho_e  &= f \big( \rho - \rho_0\big),
\\
\frac{ K - K_e}{\mu + K_e} &= f\, \bigg(\frac{ K - K_0}{\mu + K_0} \bigg) ,
\label{4b}
\\
\frac{ \mu - \mu_e}{ \mu +  \mu_e \big(1 + \frac{2\mu}K\big)}
&=
\frac{ f( \mu - \mu_0)}{ \mu + \mu_0 \big(1 + \frac{2\mu}K\big)} 
\label{4c}
\eal
\ese
where $f = r_0^2/r_1^2$. 
The relation \eqref{4a} for densities is obviously correct and therefore we will not consider density further. The expression \eqref{4b} is, as we will see,  the correct relation between $K$, $K_e$, $K_0$ and $\mu$.  However, the shell shear modulus  $\mu$  given by \eqref{4c} is not the right value, but an approximation.   
The identities \eqref{4b}, \eqref{4c}  coincide with the Hashin-Shtrikman two dimensional bounds for $K_e$ and $\mu_e$ \cite{Par-15}, similar to the three dimensional Kuster-Toks\"oz model
\cite{Berryman80}; formulae valid in the limit of small $f$ were derived in \cite[eqs.\ (3.15), (3.16)]{Kantor1982}  which  are in agreement with \eqref{4c}.
Solving for the layer or cloak properties yields
\bse{-5}
\bal{5a}
\rho    &  = \frac{\rho_e - f\rho_0}{1-f},
\\
K  &= \frac{ \frac{K_e}{\mu + K_e} - \frac{fK_0}{\mu + K_0} }
{ \frac{1}{\mu + K_e} - \frac{f }{\mu + K_0} }
\label{5b}
\eal
where $\mu$ solves the cubic equation
\bal{5c}
&\big[
(1+f)(\mu_0-\mu_e)K_eK_0 - \big( K_e+2K_0 - f(2K_e+K_0) \big) \mu_e\mu_0
\big]\,  \mu
\notag \\
+&\big[
(\mu_0-\mu_e)(K_e+fK_0) +
(1-f)(K_eK_0-2\mu_e\mu_0) + 2\mu_0K_0- f2\mu_eK_e
\big] \, \mu^2
\notag \\
+& \big[
K_e+2 \mu_0  -f (K_0 +2\mu_e) \big]\,  \mu^3
-  (1-f) \mu_0\mu_eK_0K_e = 0.
\eal
\ese
A solution  exists for ($\rho$, $K$, $\mu$)
for any given
($\rho_0$, $K_0$, $\mu_0$), ($\rho_e$, $K_e$, $\mu_e$) if $f$ is small.  As $f$ is increased, the solution may or may not exist. If $\rho_0 >\rho_e$, then
a positive solution for $\rho$ is only possible for $f < \rho_e /\rho_0$.   A small cloak is equivalent to large $f$, i.e.\ $1-f \ll 1$.

For instance, in the limiting cases when  the core is a hole,  eqs.\ \eqref{5b} and \eqref{5c} give
\beq{6-7}
K= \frac{(1+2f) \mu_e}{1 -2 \nu_e  (1+f)) },
\ \
\mu = \frac{(1+2f)\mu_e}{1-2f(1-2 \nu_e)},
\ \
 \text{for} \ K_0=\mu_0 = 0,
\eeq
where $\nu_e = \frac 12 - \frac{\mu_e}{2K_e}$. {Note that because the planar problem is isotropic, $\nu_e$ is given by the expression for the isotropic Poisson's ratio in terms of in-plane properties $K$ and $\mu$, but since the effective medium is transversely isotropic it cannot be thought of as the Poisson's ratio:  see the erratum follow up \cite{christensen1986erratum} to \cite{Christensen79}.} 
If the core is   a rigid inclusion the cloaking layer becomes
\bal{6-8}
\mu &= \frac{(2-f)\mu_e -(1+f)K_e}{2(2+f)}
+\sqrt{ \Big(\frac{(2-f)\mu_e -(1+f)K_e}{2(2+f)}\Big)^2
+\Big( \frac{1-f}{2+f}\Big) K_e\mu_e
},
\notag \\ 
K &= (1-f)K_e - f\mu ,  \ 
\ \ \text{for} \ \frac 1{K_0}=\frac 1{\mu_0} = 0.
\eal
In each case the cloak depends  on the matrix properties  and   the core volume fraction $f$.   The expression for $K$ in \eqref{6-7}, which must be positive and finite, implies that the range of possible $f$ shrinks to zero as $\nu_e$ approaches $1/2$, the incompressibility limit.

{Solutions for static cloak properties (or NIs) are now sought that 
do not require approximate effective property expressions.}



\section{Quasistatic cloaking problem setup}  \label{sec3}

 The objective is to determine necessary and sufficient conditions on the material properties of the coating of Fig.\ \ref{fig1} in order that the  combined core and coating acts as a quasi-static cloaking device.
Two distinct definitions of quasistatic cloaking are considered: (i) the neutral inclusion effect, and (ii) low frequency wave transparency. The former is a purely static concept whereby an arbitrary applied static field is unperturbed in the exterior of the core-shell composite.  Low frequency wave transparency is a dynamic concept; it requires that the leading order term in the expansion of the scattered field expressed as an expansion in frequency vanishes for  any type of  incident time harmonic plane wave. However, as one might expect, it is possible to rephrase the condition in terms of static quantities, as in Rayleigh scattering \cite{Dassios00}.  This idea is used here also, and in the process the similarities and differences between (i) and (ii) will become apparent.

Anticipating the need to go beyond isotropic shells we consider cylindrically anisotropic inhomogeneous materials \cite{Lekhnitskii} with, in general, four radially varying elastic moduli.   Our method of solution uses the  formulation of \cite{Shuvalov03withnote} although we note that other equivalent state-space approaches have been successfully employed, e.g.\ \cite{Tsukrov2010} derived  displacement and stress solutions for a multilayered 
composite cylinder with cylindrically orthotropic layers subject to homogeneous boundary loadings using the state space formalism of \cite{Tarn2001,Tarn2002}.  
Our solution method is based on  impedance matrices \cite{Shuvalov03withnote}  which do not require pointwise solutions for displacement and stress, which simplifies the analysis considerably. 

 \subsection{Matricant and impedance matrices}

{Given an arbitrary static loading in the far-field, displacement solutions may be written  in terms of summations over azimuthal modal dependence of the form $e^{in\theta}$ for integer $n$. Cylindrical coordinates $r,\theta$ are used here. Radially dependent displacements are then $u_r(r)$, $u_\theta (r)$ with associated traction components $t_r(r)$ $(=\sigma_{rr})$ and $t_\theta(r)$ $(=\sigma_{r\theta})$. Assume that the coating (cloak) is cylindrically anisotropic \cite{Shuvalov03withnote} with local orthotropic in-plane anisotropy defined by the moduli (in Voigt notation) $C_{11} $, $C_{22} $, $C_{12} $, $C_{66} $, where $1,2 \leftrightarrow r,\theta$. } The static elastic equilibrium and constitutive equations can  be written as a system of four ordinary differential equations in $r$,
\beq{1}
\frac{\dd {\bf v}}{\dd r}(r)
= 
\frac 1r
{\bf G}(r) {\bf v} (r)
\ \ \text{where}
\eeq
\beq{2}
{\bf v} = 
\begin{pmatrix}
u_r \\ -\i u_\theta \\ rt_r \\ -\i rt_\theta
\end{pmatrix},
\ \
{\bf G} =
\begin{pmatrix}
1-\gamma  &  n(\gamma-1)  & C_{11}^{-1} &0
\\
-n & 1 & 0 & C_{66}^{-1}
\\
C & -n C & \gamma -1  & n
\\
-n C& n^2 C & n (1-\gamma) &-1
\end{pmatrix},
\ \ \begin{aligned}
\gamma &= 1+ \frac{C_{12}}{C_{11}},
\\
C &= C_{22}- \frac{C_{12}^2}{C_{11}} .
\end{aligned}
\eeq
The constraint of positive definite strain energy for the two-dimensional deformation requires
$C_{11}>0$, $C_{66}>0$, $C >0$.

The propagator, or matricant, ${\bf M}(r,r_0)$, by  definition \cite{Shuvalov03withnote} satisfies
\beq{25}
\frac{\dd {\bf M}}{\dd r}
=
\frac 1r
{\bf G}(r) {\bf M}
\ \ \text{with }
\ \
{\bf M} (r_0,r_0)= {\bf I},
\eeq
where ${\bf I}$ is the $4\times 4$ identity. Note its important property that
\beq{21a}
\begin{pmatrix}
u_r(r_1) \\ u_{\theta}(r_1) \\  i r_1 t_r(r_1) \\ i r_1 t_\theta(r_1)
\end{pmatrix} =
{\bf M}(r_1, r_0)
\begin{pmatrix}
u_r(r_0) \\ u_{\theta}(r_0) \\  i r_0 t_r(r_0) \\ i r_0 t_\theta(r_0)
\end{pmatrix} .
\eeq
The 2$\times$2 impedance matrix, $ {\bf Z} (r)$,  is defined  by
\beq{21}
\begin{pmatrix}
  rt_r \\ -\i rt_\theta
\end{pmatrix} =
{\bf Z}
\begin{pmatrix}
u_r \\ -\i u_\theta
\end{pmatrix} .
\eeq
 It can  be expressed in terms of  the impedance at $r=r_0$, ${\bf Z}(r_0)$, using the
matricant, as
\beq{26}
{\bf Z}(r) = \big( {\bf M}_3+{\bf M}_4 {\bf Z}(r_0)
\big)    \big( {\bf M}_1+{\bf M}_2 {\bf Z}(r_0)
\big)^{-1}
\ \
\ \ \text{where }
\ \
{\bf M} (r,r_0)=
\begin{pmatrix}
{\bf M}_1  &  {\bf M}_2
\\
{\bf M}_3 & {\bf M}_4
\end{pmatrix}  .
\eeq
Alternatively, the impedance satisfies a separate ordinary differential matrix Riccati equation
\beq{22}
\begin{aligned}
&r \frac{\dd {\bf Z}}{\dd r}
+ {\bf Z} {\bf G}_1 +{\bf G}_1^T {\bf Z} + {\bf Z} {\bf G}_2 {\bf Z}
- {\bf G}_3 = 0 ,
\\
 \text{where} \ \
{\bf G}_1=&
\begin{pmatrix}
1-\gamma &  n (\gamma-1)
\\
-n & 1
\end{pmatrix},
\ \
{\bf G}_2= \begin{pmatrix}
  C_{11}^{-1} &0
\\
 0 & C_{66}^{-1}
\end{pmatrix},
\ \
{\bf G}_3 = C\begin{pmatrix}
1 & -n
\\
-n & n^2
\end{pmatrix}
.
\end{aligned}
\eeq
The transpose ${\bf Z}^T$ satisfies the same equation, and therefore, if the initial condition
for the impedance matrix is symmetric then it remains symmetric.  We will only consider this case, and can therefore assume that it is always symmetric, ${\bf Z}(r) = {\bf Z}^T(r)$.
The impedance matrix considered here is the static limit of the dynamic impedance discussed in  \cite{Norris10} for general cylindrical anisotropy, specifically the impedance ${\bf z}$ of \cite{Norris10} is related to the present version by ${\bf z}= - {\bf J}{\bf Z} {\bf J}^\dagger$ where ${\bf J}= \diag (1, \i)$ and $\dagger$ denotes the Hermitian transpose.   Integration of the Riccati equation for the time harmonic problem can be tricky because of the appearance of dynamic resonances, although these difficulties can be circumvented \cite{Norris2013}.   No such problems arise in the present case, for which  numerical  integration of \eqref{22} is stable.  The initial value of the impedance for a uniform cylinder is  analyzed in detail in \cite{Norris10}, where it is termed the \textit{central impedance} since it is the pointwise value of the impedance at $r=0$ required for the initial condition of the  dynamic Riccati differential equation.

The  eigenvalues of ${\bf G}$ 
are taken to be $\{\lambda_1, \lambda_2, \lambda_3, \lambda_4\}$
with right and left  eigenvectors  ${\bf v}_i$, ${\bf u}_i$ $(i=1,2,3,4)$
satisfying ${\bf G}{\bf v}_i ={\bf v}_i \lambda_i $, ${\bf u}_i^T{\bf G} =\lambda_i {\bf u}_i^T$
where ${\bf V} =[{\bf v}_1, {\bf v}_2, {\bf v}_3, {\bf v}_4]$,
${\bf U} =[{\bf u}_1, {\bf u}_2, {\bf u}_3, {\bf u}_4]$.
The eigenvectors are normalised such that
\beq{1.22}
\begin{aligned}
&{\bf U}^T{\bf V} = {\bf V}{\bf U}^T = {\bf I},
\\
&{\bf G} = {\bf V}  {\bf D} {\bf U}^T  \ \ \Rightarrow \ \
{\bf G}^m = {\bf V}  {\bf D}^m {\bf U}^T,
\\ \text{where} \ {\bf D} &= \text{diag}(
\lambda_1, \,\lambda_2,\, \lambda_3, \,\lambda_4
) .
\end{aligned}
\eeq
The eigenvalues and eigenvectors are functions of $r$ if the moduli, through ${\bf G}$ depend on $r$.

\subsubsection{Uniform properties}

For a constant set of moduli ${\bf G}$ over some range including  $r$ and $r_0$, the  eigenvalues and eigenvectors are fixed and
the solution of \eqref{25}
can be written
\beq{1.33}
{\bf M}(r,r_0) = {\bf V}  {\bf E} (r,r_0) {\bf U}^T  
\ \ \text{where} \
{\bf E} (r,r_0) = \text{diag}\big(
\big(\frac r{r_0}\big)^{\lambda_1}, \, \big(\frac r{r_0}\big)^{\lambda_2}, \,
\big(\frac r{r_0}\big)^{\lambda_3}, \, \big(\frac r{r_0}\big)^{\lambda_4}
\big) .
\eeq
Alternatively, ${\bf M}$ can be expressed simply as a matrix exponential,
\beq{1.331}
{\bf M}(r,r_0) = \de^{{\bf G}\log (r/r_0)}  = \Big(\frac r{r_0}\Big)^{\bf G} .
\eeq

There are two distinct types of impedance matrix solutions for a uniform medium.  The first is the impedance of a solid cylindrical region of finite radius.  Since there is no length scale in the  impedance relation, it follows that  the impedance is independent of the radius, and thus by virtue of eq.\ \eqref{22},  is a solution of the Riccati matrix equation
\beq{221}
{\bf Z} {\bf G}_2 {\bf Z} +
{\bf Z} {\bf G}_1 +{\bf G}_1^T {\bf Z}
- {\bf G}_3 = 0 .
\eeq
The second type of impedance is associated with the dual configuration of an infinite medium with a circular hole of finite radius.  Again, the impedance is a root of \eqref{221}.  These  matrix roots of the algebraic Riccati equation can be found using standard matrix algorithms \cite{Higham:2008:FM,Norris2013a}.

\subsection{Long wavelength scattering}

{The exterior medium is strictly transversely isotropic but we are interested in planar wave propagation. Hence this two-dimensional problem is isotropic with mechanical behaviour characterised by the two elastic properties $\mu_e$ and $K_e=\lambda_e+\mu_e$. The displacement can thus be expressed using two potential functions $\phi$ and $\psi$},
\beq{-1}
{\bf u} = \nabla\phi - \nabla \times \psi {\bf e}_3 .
\eeq
Assuming time dependence $e^{-\i\omega t}$, the incident wave is
in the $x_1$-direction
$\phi = A_L \de^{\i k_L x_1}$,  $\psi=A_T \de^{\i k_T x_1}$,
where $k_L=\omega /c_L$, $k_T=\omega /c_T$, $c_L^2={(K_e+\mu_e)}/\rho$,
$c_T^2={\mu_e}/\rho$
and $A_L$, $A_T$ are the longitudinal and  transverse wave amplitudes.
Taking $A_L= (\i k_L)^{-1}$, $A_T= (\i k_T)^{-1}$ leads to the incident wave
\beq{-12}
{\bf u}
= \big(  \de^{\i k_L x_1},\,  \de^{\i k_T x_1} ,\, 0\big)  ,
\ \
{\bf v} = {\bf v}_L + {\bf v}_T
\eeq
where,  using $x_1=r\cos\theta$, $x_2=r\sin\theta$, 
\beq{-13}
{\bf v}_L
= \begin{pmatrix}
\cos\theta
\\
\i \sin\theta
\\
\i k_L r({\lambda_e} + 2{\mu_e} \cos^2\theta )
\\
-k_L r  2{\mu_e} \cos\theta \sin\theta
 \end{pmatrix}
e^{\i k_L r\cos\theta} ,
\ \
{\bf v}_T
= \begin{pmatrix}
\sin\theta
\\
-\i \cos\theta
\\
\i k_T r 2{\mu_e} \cos\theta \sin\theta
\\
k_T r  {\mu_e} (\cos^2\theta -\sin^2\theta)
 \end{pmatrix}
e^{\i k_T r\cos\theta} .
\eeq
In the low-frequency, or equivalently long-wavelength regime, and in the vicinity of the cylinder,
\beq{00}
k_L  r \ll 1, \ \ k_T  r \ll 1 ,
\eeq
 resulting in the asymptotic expansions
\bse{-131}
\bal{-13a}
{\bf v}_L
&=
\frac {\i}2 k_L  r \begin{pmatrix}
1
\\
0
\\
2({\lambda_e} + {\mu_e}  )
\\
0
 \end{pmatrix}
+ \begin{pmatrix}
\cos\theta
\\
\i \sin\theta
\\
0 \\
0 \end{pmatrix}
+\frac {\i}2 k_L  r \begin{pmatrix}
\cos 2\theta
\\
\i \sin 2\theta
\\
  2{\mu_e} \cos 2\theta
\\
\i  2{\mu_e} \sin 2\theta
 \end{pmatrix}
+\text{O} \big( k_L^2r^2\big),
\\   
{\bf v}_T
&= \frac {1}2 k_T  r
\begin{pmatrix}
0
\\
1
\\
0
\\
0
 \end{pmatrix}
+ \begin{pmatrix}
\sin\theta
\\
-\i \cos\theta
\\
0 \\
0 \end{pmatrix}
+\frac {\i}2 k_T  r \begin{pmatrix}
\sin 2\theta
\\
-\i \cos 2\theta
\\
  2{\mu_e} \sin 2\theta
\\
-\i  2{\mu_e} \cos 2\theta
 \end{pmatrix}
+\text{O} \big( k_T^2r^2\big) .
\eal
\ese
These can be considered as {\it near-field} expansions, valid in the neighborhood for which
\eqref{00} holds.
The first term in ${\bf v}_T$ is a rigid body rotation, and the second terms in both ${\bf v}_L$ and ${\bf v}_T$ are rigid body translations.
The first term in ${\bf v}_L$ can be interpreted as  a radially symmetric far-field loading, while the {third} terms in both ${\bf v}_L$ and ${\bf v}_T$ are $n=\pm 2$ shear-type loadings. The $n=1$ loadings cause the inclusion to undergo rigid body motion;  the  parameter that is relevant in the low frequency limit is the effective mass, or equivalently its effective density.  Therefore, at this level of long-wavelength approximation, the scattering can be evaluated from the solutions for $n=0$ and $n=\pm 2$ quasi-static loadings.  In order to better identify these terms, we rewrite \eqref{-131} as
\bse{-132}
\bal{-132a}
{\bf v}_L
&=
\frac {\i}2 k_L  r \begin{pmatrix}
1
\\
0
\\
2({\lambda_e} + {\mu_e}  )
\\
0
 \end{pmatrix}
+
\sum_{n=\pm 1} \frac{e^{\i n\theta}}2
\begin{pmatrix}
{\bf a}_\pm
\\
0 \\
0 \end{pmatrix}
+\frac {\i}4 k_L  r
\sum_{n=\pm 2} \de^{\i n\theta}
\begin{pmatrix}
{\bf a}_\pm
\\
  2{\mu_e} {\bf a}_\pm
 \end{pmatrix}
+\text{O} \big( k_L^2r^2\big),
\\   
\i {\bf v}_T
&= \frac {\i}2 k_T  r
\begin{pmatrix}
0
\\
1
\\
0
\\
0
 \end{pmatrix}
+ \sum_{n=\pm 1} \frac{e^{\i n\theta}}2
\begin{pmatrix}
\pm {\bf a}_\pm
\\
0 \\
0 \end{pmatrix}
+\frac {\i}4 k_T  r \sum_{n=\pm 2}\de^{\i n\theta}
\begin{pmatrix}
\pm {\bf a}_\pm
\\
 \pm  2{\mu_e} {\bf a}_\pm
 \end{pmatrix}
+\text{O} \big( k_T^2r^2\big)
\eal
\ese
where ${\bf a}_\pm \equiv
\begin{pmatrix}
1 \\ \pm 1
 \end{pmatrix}
$.
The terms in these equations for the incident plane waves can be identified as separate quasistatic loadings of type $n=0$, $1$, and $2$.  The $n=1$ term involves only the effective mass term, which involves  the average density. {This is decoupled from the elasticity problem} and will not be discussed further.  For the remainder of the paper we will focus on the $n=0$ and $n=2$ loadings.

\section{Effective bulk modulus: $n=0$}  \label{sec4}

Equations \eqref{1} and \eqref{2} simplify for $n=0$ to two uncoupled systems
\bal{3}
\frac{\dd  }{\dd r}
\begin{pmatrix}
u_r \\ rt_r \\
\end{pmatrix}
= &
\frac 1r
\begin{pmatrix}
1-\gamma   & C_{11}^{-1}  ,
\\
C &  \gamma -1
\end{pmatrix}
 \begin{pmatrix}
u_r \\ rt_r \\
\end{pmatrix} ,
\\ \frac{\dd  }{\dd r}
\begin{pmatrix}
 u_\theta \\ rt_\theta
\end{pmatrix}
 = &
\frac 1r
\begin{pmatrix}
1 &  C_{66}^{-1}
\\
0 &-1
\end{pmatrix}
\begin{pmatrix}
 u_\theta \\ rt_\theta
\end{pmatrix} .
\label{4}
\eal
The latter is associated with pure twist or torsion: define the relation between  the angle of twist
and the angular traction  as $r^{-1} u_\theta=S_e(r) t_\theta$ then eq.\ \eqref{4} implies that the
effective compliance is
\beq{--4}
S_e(r)  = \frac{r^2}{r_0^2} S_0 + r^2 \int_{r_0}^r \frac{\dd x}{x^3 C_{66}(x)}
\eeq
where $S_0 = S_e(r_0)$.
For instance, $S_0=0$ for a shell $r>r_0$ pinned at $r=r_0$.

Our main concern with the $n=0$ case  is  for
 radially symmetric loading for which the only quantity of importance is the effective compressibility of the inclusion. Define the pointwise effective bulk modulus $K_*$ as a function of $r$,  by
\beq{5}
K_* (r) \equiv \frac{rt_r}{2u_r} .
\eeq
Matching this to the exterior medium guarantees a neutral inclusion effect for $n=0$, in addition to zero monopole scattering in the low frequency regime.  We next derive
$K_* (r)$.

\subsection{A scalar Riccati equation for the bulk modulus}

Substituting $rt_r = 2K_* u_r$ in \eqref{3} yields  the Riccati ordinary differential equation
\beq{6}
\frac{\dd K_* }{\dd r}
+\frac 2r C_{11}^{-1}
\Big(
K_*^2 -C_{12}K_* -\frac 14 (C_{11}C_{22}-C_{12}^2)
\Big)=0.
\eeq
Noting that $C_{11}C_{22}-C_{12}^2 >0$, define the positive moduli
$K$, $\mu$ and the non-dimensional parameter $\beta >0$
\beq{7}
K = \frac12 \big(\sqrt{C_{11}C_{22}} +C_{12} \big),
\quad 
\mu = \frac12 \big(\sqrt{C_{11}C_{22}} -C_{12} \big),
\quad 
\beta = \sqrt{C_{22}/C_{11}} ,
\eeq
so that the Riccati equation  becomes
\beq{8}
\frac{\dd K_* }{\dd r}
+\frac 2r \beta (K+\mu )^{-1} \, \big(K_*  -K\big)\big(K_*  +\mu\big)=0.
\eeq

\subsection{Example: Constant moduli}
If $K$, $\mu$ and $\beta$  are constants the Riccati equation  \eqref{8} can be integrated and combined 
with the matching conditions at the core boundary,  $K_*(r_0)=K_0$, and at the exterior boundary,  $K_e=K_*(r_1)$, $r_1\ge r_0$,  to yield
\beq{10}
\frac{ K - K_e}{\mu + K_e} = f^\beta\, \bigg(\frac{ K - K_0}{\mu + K_0} \bigg)
\ \ \text{where} \ f = \frac{r_0^2}{r_1^2}.
\eeq 
This is in agreement with \eqref{4b} when 
$\beta = 1$, but is more general in that it includes the possibility of an anisotropic layer  $(\beta \ne 1)$. 
For given values of the inner and outer parameters $K_0$, $K_e$ and radii $r_0$, $r_1$,  the relation \eqref{10} places a constraint on the possible cloaking moduli.   In this case, it relates $K$, $\mu$ and $\beta$ according to
\beq{12}
\mu = - \bigg(
\frac
{ K_0 (K_e - K)    - K_e(K_0 - K)  f^\beta  }
{     (K_e - K)    -    (K_0 - K)  f^\beta  }
\bigg) .
\eeq
For instance, taking $K_0 \to \infty$, $0$, implies the limiting cases
\beq{13}
\mu = \begin{cases}
(K_e - K)f^{-\beta}    -    K_e   
, &
\text{rigid core},
\\ 
\Big( (K_e^{-1} - K^{-1})f^{-\beta}    -    K_e^{-1}\Big)^{-1} , &
\text{hole} .
\end{cases}
\eeq

At this stage there are still two unknowns, $K$ and $\mu$, and only one relation between them, Eq.\ \eqref{12}. {Choosing coatings with in-plane shear and bulk moduli and anisotropy ratio $\beta$ that satisfy the relationship \eqref{12} thus ensures a neutral inclusion when the medium is subjected to in-plane hydrostatic pressure. In order to find a second relationship between $\mu$ and $K$ (and thus uniquely define the coating properties) a second relation needs to be determined, if one exists. It transpires that this second relationship comes from the $n=2$ solution.}

\section{Effective shear modulus: $n=2$} \label{sec5}

We  first consider the cloaking layer to be isotropic, and prove that it is not possible to obtain a strong neutral inclusion (static cloak).  We will then show that the strong NI condition can only be met with an anisotropic layer. 

\subsection{Isotropic  medium}
An isotropic shell has two elastic parameters which can be taken as $C_{66}$ and $\gamma$, in terms of which  the remaining two elastic moduli in Eq.\ \eqref{2} are
$C_{11}= {2 C_{66}}(2- \gamma )^{-1}$ and $C= 2 C_{66} \gamma$.
The eigenvalues of ${\bf G}$ are $n-1$, $n+1$, $1-n$, $-1-n$, which for $n=2$ become $\{\lambda_1, \lambda_2, \lambda_3, \lambda_4\}= \{1,3,-1,-3\}$.  The  right and left  eigenvectors satisfying \eqref{1.22}  are
\beq{1.2}
{\bf U} =  \frac 12 {\bf C}
\begin{bmatrix}
 2\gamma& - 3 &  1 & 0 \\ & &&\\
 -  \gamma &  3 &  1 & - 3  \gamma \\ &&& \\
 2& - 1 & - 1 & 2\gamma -2 \\&&& \\
 {2 - \gamma}&  1 & - 1 & {2 + \gamma}
\end{bmatrix},
\ \
{\bf V} =  \frac 12 {\bf C}^{-1}
\begin{bmatrix}
1 & \frac {2\gamma -2}3& 2& \frac 13 \\&&& \\
1  & \frac{2 + \gamma}3& {2 - \gamma}& -\frac 13 \\&&& \\
 1& 0& -2  \gamma & -1 \\&&& \\
 1& \gamma& \gamma& 1
\end{bmatrix}
\eeq 
where ${\bf C} = \text{diag}( 2C_{66}, \, 2C_{66}, \, 1,\, 1)$, $\gamma = \frac 1{1-\nu}$ 
and $\nu$ is Poisson's ratio.

Consider a solid cylinder.  Only   solutions with $\lambda_i \ge 0$ are permissible in the cylinder, corresponding to the first two columns of ${\bf V}$ in \eqref{1.2}.  The impedance matrix at every point in the cylinder is  then constant and equal to
\beq{27}
{\bf Z}  = {\bf V}_3{\bf V}_1^{-1}
 = \frac  {2C_{66}}{4-\gamma }
\begin{pmatrix}
2+\gamma & 2-2\gamma
\\
2-2\gamma & 2+\gamma
\end{pmatrix}
 \ \ \text{where}\ \
{\bf V}  =
\begin{pmatrix}
{\bf V}_1  &  {\bf V}_2
\\
{\bf V}_3 & {\bf V}_4
\end{pmatrix} ,
\eeq
in agreement with  \cite[eq.\ (8.6)]{Norris10} for the central impedance matrix.
It may be checked that  ${\bf Z}$ of \eqref{27} solves the Riccati equation \eqref{221}.

\subsection{Neutral inclusion {shear} condition}

A cylinder of uniform material with shear modulus and Poisson's ratio $\mu_0$, $\nu_0$ and radius $r_0$ is surrounded by a shell, or cloak with outer radius $r_1$.  The impedance on the exterior boundary is, see \eqref{26}
\beq{261}
{\bf Z}(r_1) = \big( {\bf M}_3+{\bf M}_4 {\bf Z}(r_0)
\big)    \big( {\bf M}_1+{\bf M}_2 {\bf Z}(r_0)
\big)^{-1},
\ \
{\bf Z} (r_0)= \frac {2\mu_0}{3-4\nu_0 }
\begin{pmatrix}
3-2\nu_0 &  -2\nu_0
\\
-2\nu_0 & 3-2\nu_0
\end{pmatrix}
\eeq
and ${\bf M}_i$ are block elements of the matricant ${\bf M} (r_1,r_0)$.   The far-field loading  for $n=2$  ($n=-2$ is different!) follows from
\eqref{-132}.  In addition, the exterior field in $r>r_1$ comprises the solutions with $\lambda_i <0$ which are ${\bf v}_3$ and ${\bf v}_4$, the third and fourth columns in
${\bf V}$ of \eqref{1.2}.  The continuity condition at the interface $r=r_1$ for some incident amplitude $\alpha_1 \ne 0 $ is
\beq{31}
\alpha_1
\begin{pmatrix}
1 \\ 1 \\
2\mu_e \\ 2\mu_e
\end{pmatrix}
+\alpha_3 {\bf v}_3 +\alpha_4{\bf v}_4 =
\begin{pmatrix}
{\bf b}
\\
 {\bf Z}(r_1) {\bf b}
\end{pmatrix}
\eeq
where $\mu_e$ 
is the exterior shear modulus.     The strong neutral inclusion condition requires that
$\alpha_3=0$, $\alpha_4=0$, in which case we have
\beq{32}
{\bf b} =
\alpha_1
\begin{pmatrix}
1 \\ 1
\end{pmatrix}  ,
\ \
 {\bf Z}(r_1) {\bf b}
= 2\mu_e \alpha_1
\begin{pmatrix}
1 \\ 1
\end{pmatrix}
\ \ \text{for a strong NI}.
\eeq
{Hence, the strong neutral inclusion condition is that
$(1\ 1)^T$ is an eigenvector of $ {\bf Z}(r_1) $
with eigenvalue $2\mu_e$.  }
The first of these requires that
\beq{33}
Z_{11}+Z_{12} = Z_{21}+Z_{22},
\eeq
which  can be simplified using the fact that the impedance is symmetric, $Z_{12} = Z_{21}$.  Thus,
the cylindrical region $r\le r_1$ acts as a strong neutral inclusion  if and only if
the elements of the impedance matrix satisfy
\beq{35-}
Z_{11} (r_1) =  Z_{22} (r_1) \ \ \text{for a strong NI}.
\eeq

\subsubsection{Isotropic core plus shell}

Consider a core $\mu_0$, $\nu_0$ of radius $r_0$ with a surrounding shell
$\mu$, $\nu$ of outer radius $r_1 >r_0$.   Using eqs.\ \eqref{1.33}, \eqref{1.2} and \eqref{261} it can be shown that
\beq{51}
Z_{11} (r_1) -  Z_{22} (r_1) =
3 \big(\frac {r_1^2}{r_0^2} - 1\big) \frac{ (\mu_0 - \mu)}{ (1-\nu)\mu}
\Big( \mu + \frac{\mu_0}{3-4\nu_0} \Big) / \det  \big( {\bf M}_1(r_1,r_0)+{\bf M}_2(r_1,r_0) {\bf Z}(r_0)
\big) .
\eeq
The neutral inclusion condition \eqref{35-} can only be met if the shell and core have the same shear modulus, $\mu_1=\mu_0$, in which case the
effective shear modulus  is simply $\mu_0$, regardless of the values of the Poisson's ratios $\nu_1$ and $\nu_0$, see Eq.\  \eqref{36}.  This means that the core cannot be transformed into a neutral inclusion by surrounding it with a single shell of isotropic material.

\subsection{Low frequency transparency condition {in shear}}

For a given incident wave the scattered displacement ${\bf u}^s$  in the exterior of the inclusion
can be expressed using eq.\ \eqref{-1} with
$\phi = B_L H_2^{(1)} (k_Lr)e^{\i 2\theta}$,
$\psi = \i B_T H_2^{(1)} (k_Tr)e^{\i 2\theta}$,
where $H_n^{(1)}$ is the Hankel function of the first kind.  This yields, dropping the
$e^{i2\theta} $ term,
\beq{-39}
\begin{aligned}
u^s_r &=  k_L B_LH_2^{(1)'}(k_Lr) + 2 \frac{B_T}r H_2^{(1)}(k_Tr) ,
			\\
	-\i u^s_\theta &= 	  k_T B_TH_2^{{(1)}'} (k_Tr)  +     2 \frac{B_L}r H_2^{(1)}(k_Lr) .
\end{aligned}
\eeq
Both $B_L$ and $B_T$ are functions of frequency, the precise forms dependent on the inclusion details.  For the moment we assume that they each have regular expansions about $\omega = 0$, i.e.
\beq{=3-}
\begin{aligned}
B_L &= B_{L0} + \omega B_{L1} + \ldots ,  \\
B_T &= B_{T0} + \omega B_{T1} + \ldots  .
\end{aligned}
\eeq
Expanding \eqref{-39} in the  long wavelength near-field limit,
the scattered wave is  to leading order in $\omega$,
\beq{39}
\begin{pmatrix}
u^s_r \\
\i u^s_\theta
\end{pmatrix}
=  -\frac{2\i}{\pi r}
\begin{pmatrix}
B_{T0} \\
B_{L0}
\end{pmatrix}
+\frac{8\i}{\pi r^3}
\Big(  \frac{B_{L0}}{k_L^2}  - \frac{B_{T0}}{k_T^2}   \Big)
\begin{pmatrix}
1 \\
-1
\end{pmatrix}
 + \ldots .
\eeq
This low frequency expansion should be consistent with the purely static representation of the exterior "scattered" field as a sum of the form,  see eq.\ \eqref{31},
\beq{45=}
{\bf v}^s =
\alpha_3 {\bf v}_3 +\alpha_4{\bf v}_4
\eeq
where ${\bf v}_3 $ and ${\bf v}_4 $ are the third and fourth columns in ${\bf V}$ of \eqref{1.2}, corresponding to  $r^{-1}$ and $r^{-3}$ decay outside the inclusion, respectively.
Comparing the $r^{-1}$ term in \eqref{39} with the first two elements of ${\bf v}_3 $ implies
that
\beq{40=}
\frac{B_{L0}}{B_{T0}} = \frac {2-\gamma}2
\ \ \Rightarrow \ \      \frac{B_{L0}}{k_L^2} =  \frac{B_{T0}}{k_T^2}
\eeq
because  $1-  \frac {\gamma}2 = k_L^2 /k_T^2$.  Equation \eqref{40=} means that   if one of
$B_{L0}$, $B_{T0}$ vanishes, then both vanish. Equivalently, it says that both $B_{L0}$, $B_{T0}$ vanish if the coefficient of the $r^{-1}$ term  is zero.  

Hence, the leading order term in the low frequency expansion of the scattered field  vanishes
iff the coefficient of the $r^{-1}$ term, i.e.\ ${\bf v}_3$, in the quasi-static  solution is zero. Low frequency transparency therefore requires only that $\alpha_3$ vanishes.  This result agrees with the strain energy condition first derived by
\cite{Christensen79}, and later in more general form by \cite{Hashin1990,Herve95}.
Also,  the above derivation is independent of the type of incident wave, but relies only on the form of the scattered wave potentials as a combination of Hankel functions. 

In summary, we conclude that
\begin{lemma}
  Low frequency transparency {in shear}, is obtained if  (see Eq.\ \eqref{45=})
\beq{90}
\alpha_3 = 0,
\eeq
{which is equivalent to the weak neutral inclusion condition for in-plane deformation, if \eqref{12} also holds (the bulk modulus condition) and $\beta=1$ for isotropic coatings. }

{The core-shell composite is a strong neutral inclusion for planar deformation if and only if}
\beq{90-1}
\alpha_3 = 0 \ \text{and }\  \alpha_4 = 0 .
\eeq
{together with \eqref{12}}.
\end{lemma}

We next seek more explicit versions of these conditions, and in the process find the effective shear modulus of the matrix. 

\subsubsection{The effective shear modulus}

Equation \eqref{31} can be written
\beq{38}
\Big[ - {\bf v}_3 \ \, -{\bf v}_4 \ \ \begin{matrix} {\bf I} \\ {\bf Z}(r_1) \end{matrix}\Big]
\begin{pmatrix}
 \alpha_3 \\ \alpha_4 \\  {\bf b}
\end{pmatrix}
=
\alpha_1
\begin{pmatrix}
1 \\ 1 \\
2\mu_e \\ 2\mu_e
\end{pmatrix}  .
\eeq
The transparency/weak NI condition \eqref{90} then becomes
\beq{53}
\det
\Big[  {\bf v}_1 \ \, {\bf v}_4 \ \, \begin{matrix} {\bf I} \\ {\bf Z}(r_1) \end{matrix}\Big]
=0
\ \ \ \Rightarrow \det
\begin{pmatrix}
\frac 12  & \frac 16 & 1 & 0 \\
\frac 12  & -\frac 16 & 0 & 1  \\
\mu_e & -\mu_e & Z_{11} & Z_{12} \\
\mu_e & \mu_e & Z_{12} & Z_{22}
\end{pmatrix}
=0 .
\eeq
Expanding the determinant yields a quadratic equation for the effective shear modulus
\beq{54}
\mu_e^2 - \frac{\mu_e}6 \big( Z_{11}+Z_{22} + 4Z_{12}\big)
-\frac 1{12}\big( Z_{11}Z_{22} -Z_{12}^2\big) =0.
\eeq
The sign of the root chosen must agree with the neutral inclusion value for the effective modulus above when condition \eqref{35-} holds.

The  above results for both the low frequency transparency and the weak and strong neutral inclusion conditions can be combined {with the bulk modulus condition of Section \ref{sec4}} as follows.
\begin{thm}
The cylindrical core-shell  is transparent at low frequency and acts as a weak neutral inclusion {for in-plane hydrostatic pressure and in-plane shear} if the exterior medium has {bulk and} shear moduli
\bse{3=0}
\bal{3=5}
K_e &= \frac{K(\mu +K_0) - \mu (K-K_0)(r_0/r_1)^{2\beta }}
{\mu +K_0 + (K-K_0)(r_0/r_1)^{2\beta} }, 
\\
\mu_e &= \frac 16\bigg( Z_s+2Z_{12} + \sqrt{ (2Z_s+ Z_{12})^2 - 3Z_d^2}
\bigg)  
 \label{+1}
\eal
where 
\bal{3=6}
K &= \frac12 \big(\sqrt{C_{11}C_{22}} +C_{12} \big),
\quad 
\mu = \frac12 \big(\sqrt{C_{11}C_{22}} -C_{12} \big),
\quad 
\beta = \sqrt{C_{22}/C_{11}} ,
\\
Z_s &= \frac 12\big( Z_{11}+Z_{22}\big),
\quad 
Z_d = \frac 12\big( Z_{11}-Z_{22}\big),
\eal
\ese
and $Z_{ij}$ are the elements of the impedance matrix ${\bf Z}(r_1)$ defined by eq.\ \eqref{261}. 

Furthermore, the layered shell acts as a strong neutral inclusion {for in-plane shear and hydrostatic loading}, if and only if
\beq{35}
Z_d = 0
\eeq
coupled with \eqref{3=5}. {The strong NI condition \eqref{35}}
 cannot occur if the shell is isotropic, but requires strict anisotropy. If the strong NI condition is met then the matrix shear modulus is
\beq{36}
\mu_e = \frac 12 (
Z_{11}+Z_{12} ) .
\eeq
{and the matrix bulk modulus is $K_e$ of \eqref{3=5}.}
\end{thm}
{In practice of course the \textit{core} and \textit{matrix/exterior} properties are specified and the coating properties (and thickness) are deduced by solving \eqref{36} and \eqref{3=5}.} 

In summary, low frequency transparency/weak NI can be achieved with isotropic shell material.  The strong neutral inclusion condition restricts the types of shells: Eq.\ \eqref{51} indicates that a uniform isotropic shell cannot yield a strong NI, regardless of the isotropic core properties.
We note that the explicit form of $\mu_e$ in \eqref{+1} is far simpler than the alternatives available \cite[eq.\ (50)]{Hashin1990}, \cite[eq.\ (82)]{Herve95}, even the original \cite[eq.\ (4.11)]{Christensen79}.  Finally, it should be kept in mind that, just as for the approximate NI considered in Section \ref{sec2}, the weak and strong NI conditions are not guaranteed to be achievable for all combinations of matrix and core properties and core volume fraction in the composite cylinder. 

\section{{Implementation of the theory: weak vs. strong neutral inclusions}}\label{sec6}

{We now provide some examples of strong NIs. In particular, for each of the examples in Table \ref{-6} the identities \eqref{3=5} and \eqref{35} are satisfied, and hence the composite cylinder is a strong NI. }
\begin{table}[h]
\centering
\begin{tabular}{@{}|l|l|l|l|l|l|l|l||l|l|@{}}
\hline
Example & $r_0/r_1$ & $\mu_0$ & $\nu_0$   & $C_{11}$ & $C_{22}$ & $C_{12}$ & $C_{66}$ & $K_e$  & $\mu_e$ \\
\hline
&    &       &   &        &  &        &  &      &                         \\
(i) & 0.5      & 1       & $\frac 13$ & 4.0      & 4.40       & 2.49     & 1.0      & 3.2572 & 0.9401  \\
(ii) & 0.2        & 0       & $\frac 13$ & 4.8      & 2.9781     & 2.80     & 1.0      & 1.9782 & 0.6445   \\
(iii) & 0.2     & 10$^6$  & $\frac 13$ & 2.5782   & 2.52       & 2.4      & 1.0      & 2.5848 & 0.2990  \\
(iv) & 0.75   &  21     & $\frac 13$ & 3.5839   & 11.5096  & -0.4658    &3.2078    & 6.0307 & 5.9049  \\
\hline\hline
\end{tabular}
\caption{{Examples of composite cylinder strong NIs.  The isotropic core radius and properties are
$r_0$, $\mu_0$, $\nu_0$ and the exterior (matrix) properties are the in-plane bulk and shear moduli $K_e$ and  $\mu_e$. The anisotropic coating (shell) properties required are $C_{11}$, $C_{22}$, $C_{12}$ and $C_{66}$ and its outer radius is $r_1=1$.}
}
\label{-6}
\end{table}

\begin{figure}[htbp]
\centering
	\makebox[\textwidth]{
		\raisebox{25ex}{(a)}
		\includegraphics[width=6cm]{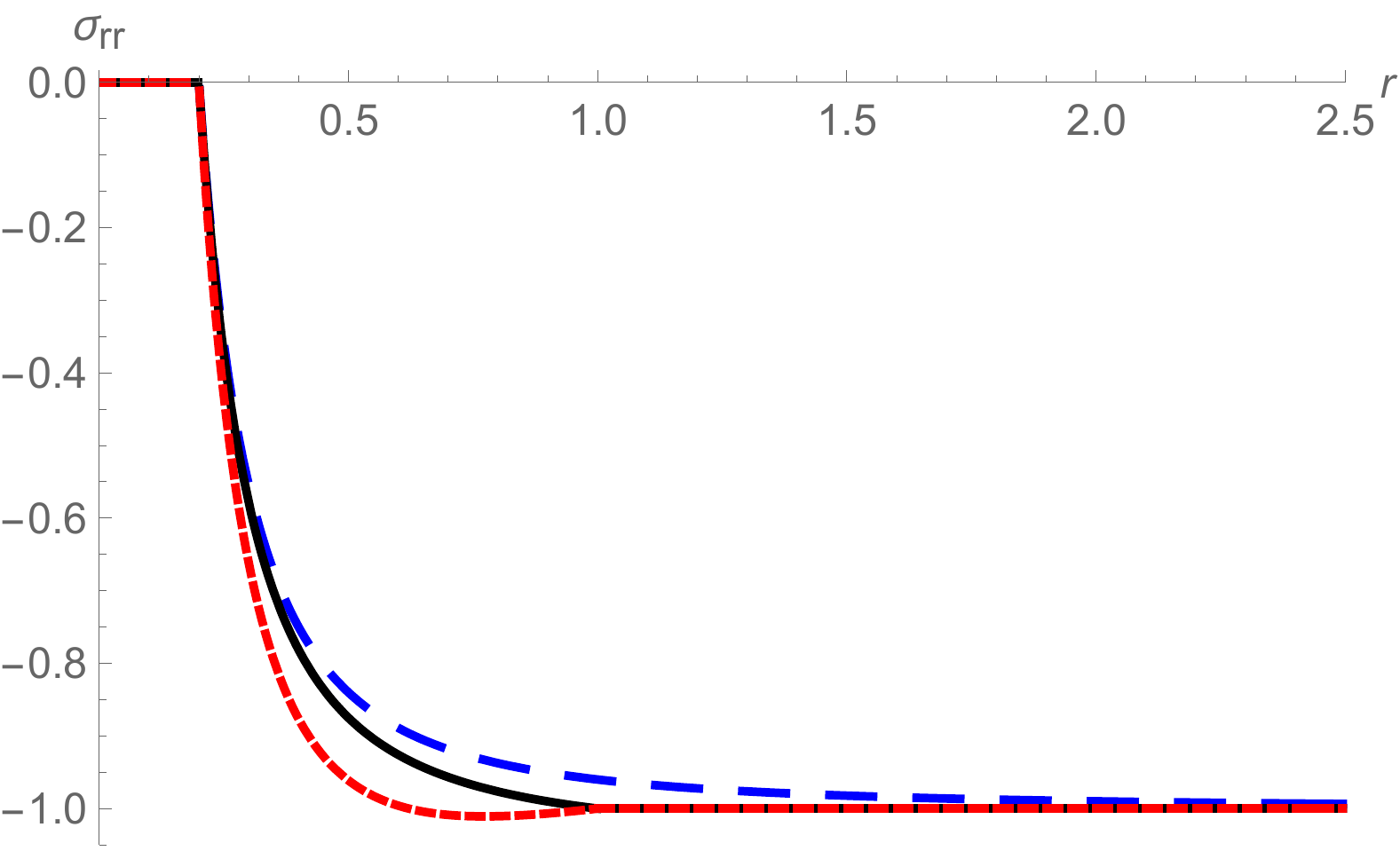}
		\hfill
		\raisebox{25ex}{(b)}
		\includegraphics[width=6cm]{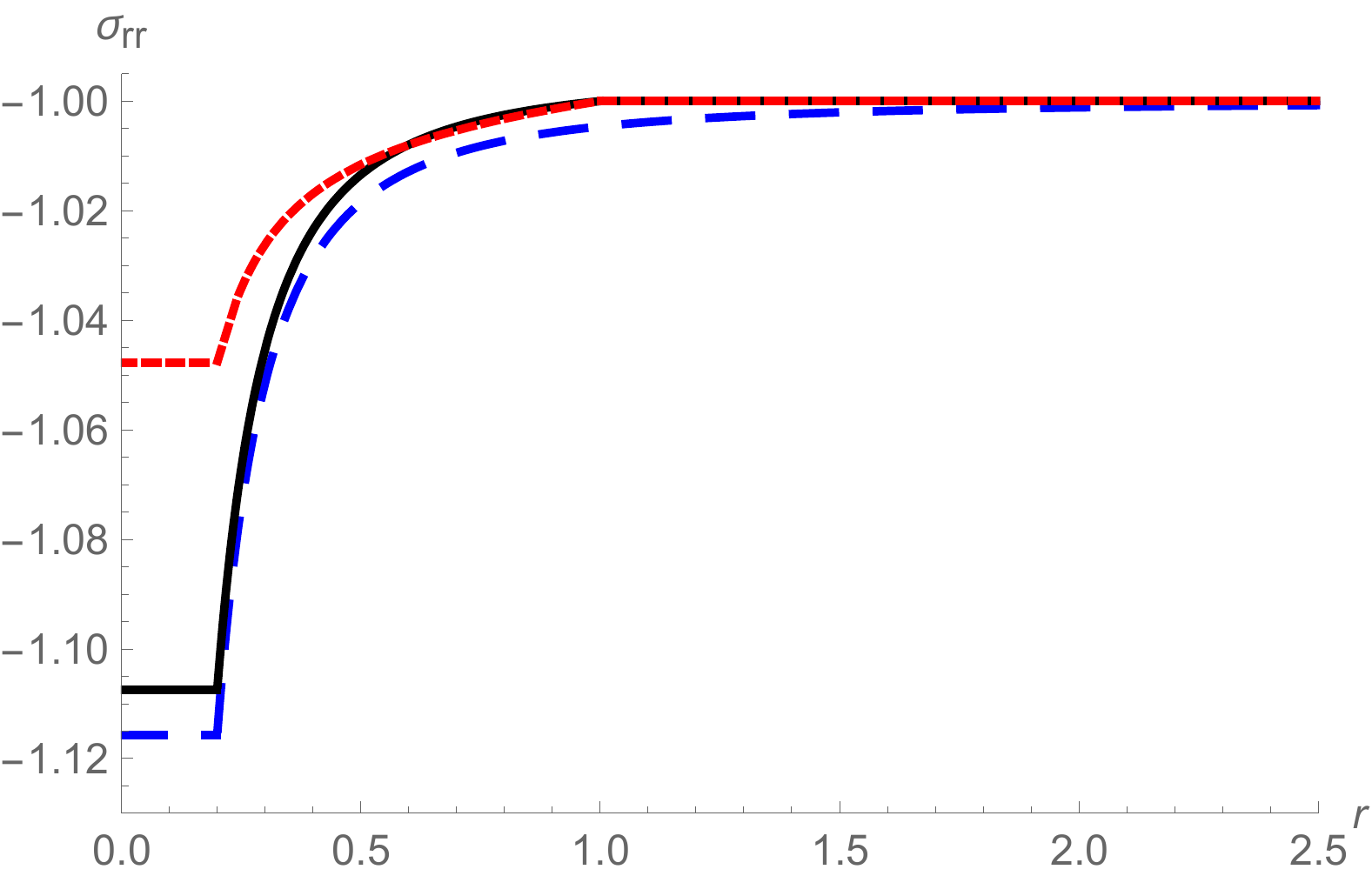}
	}
	\makebox[\textwidth]{
		\raisebox{25ex}{(c)}
		\includegraphics[width=6cm]{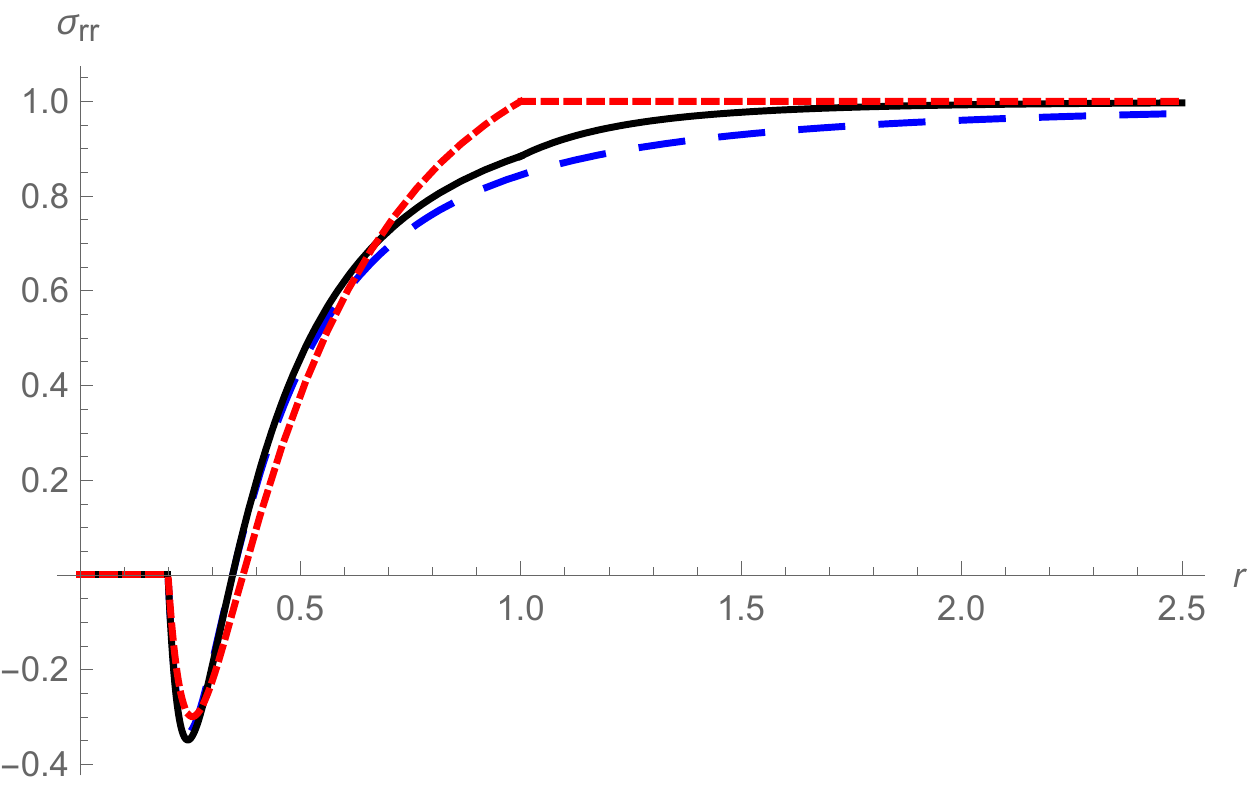}
		\hfill
		\raisebox{25ex}{(d)}
		\includegraphics[width=6cm]{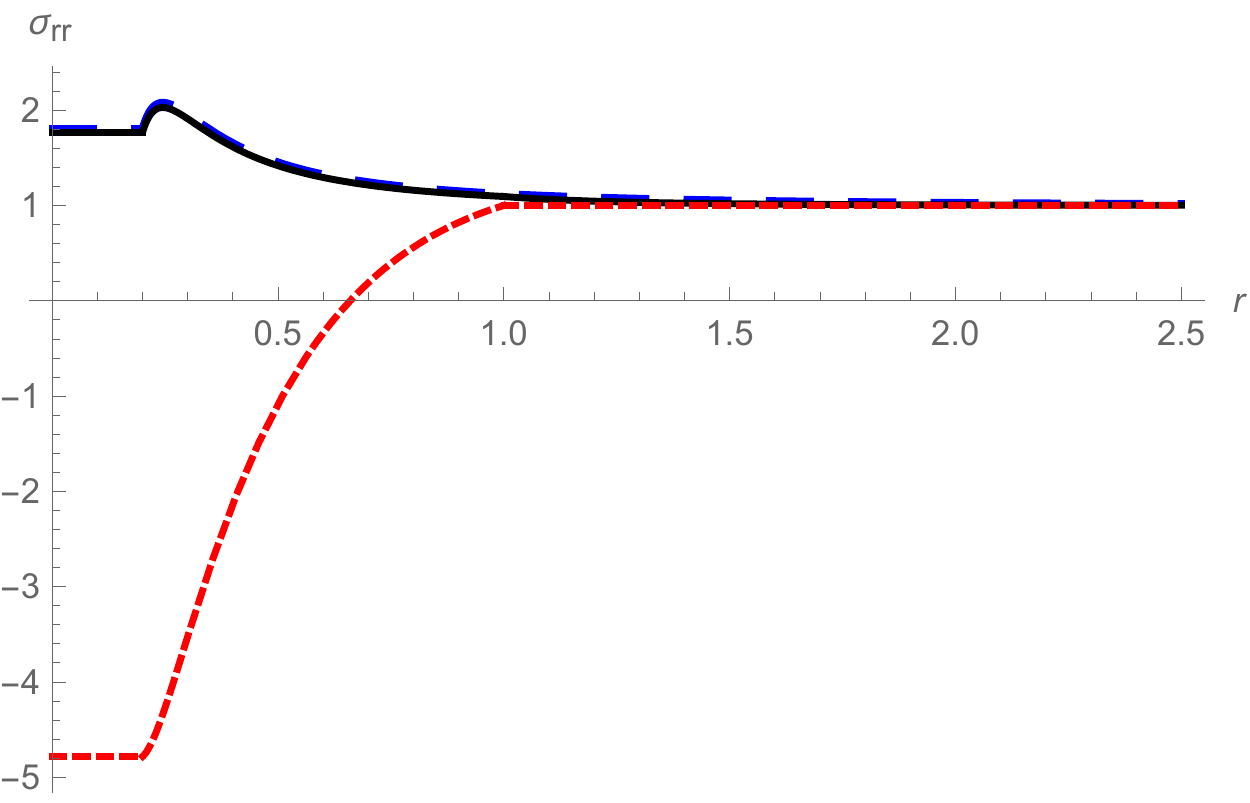}
	}
	\makebox[\textwidth]{
		\raisebox{25ex}{(e)}
		\includegraphics[width=6cm]{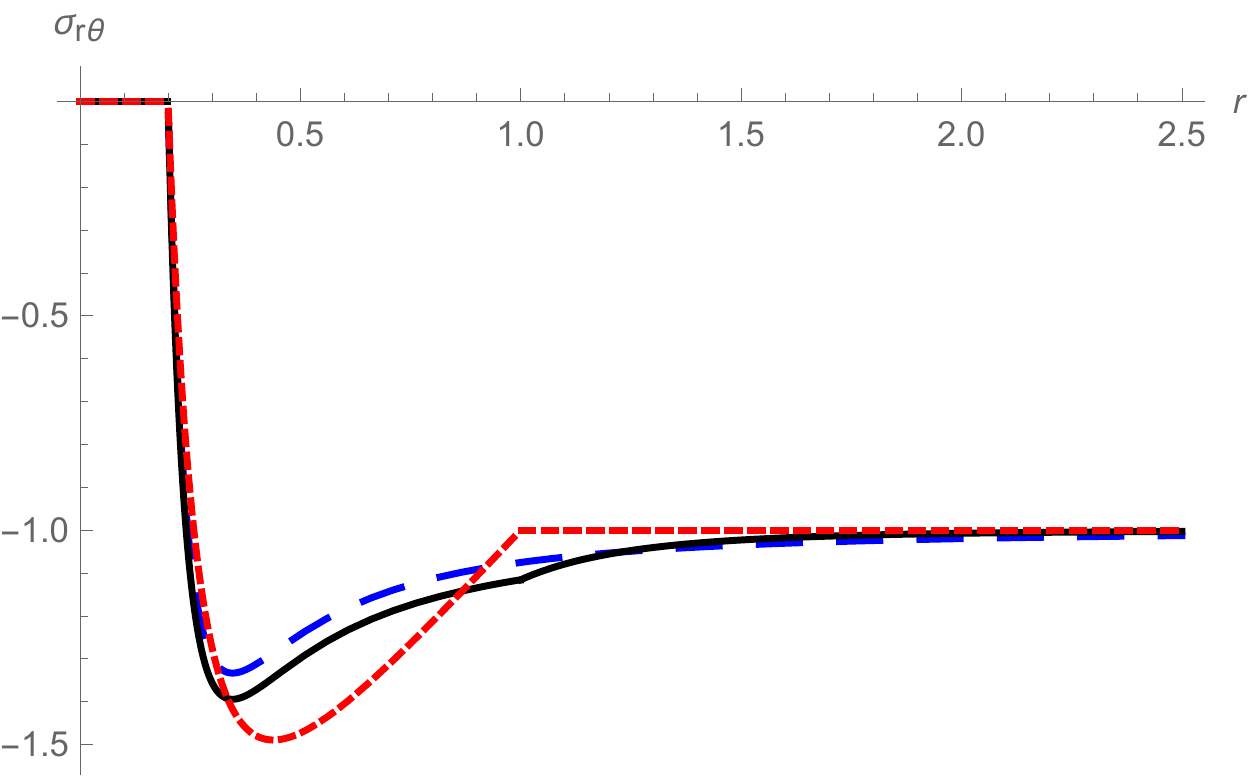}
		\hfill
		\raisebox{25ex}{(f)}
		\includegraphics[width=6cm]{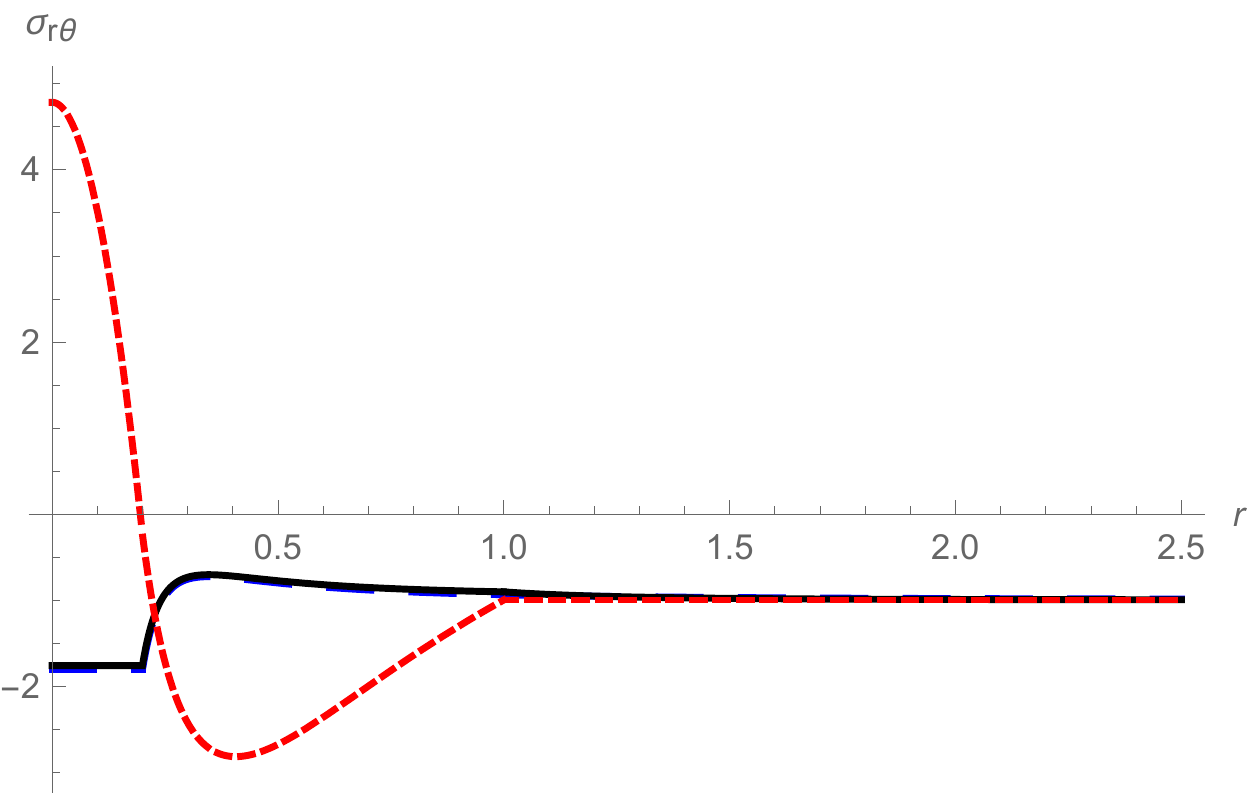}
	}
\caption{Stress plots as functions of $r$, associated with an isolated fibre (blue long dash), weak (isotropic) NI (black solid), strong (anisotropic) NI (red short dash). (a), (c), (e): Example (ii) and (b), (d), (e): Example (iii) from Table \ref{-6}.
(a) and (b) illustrate $\sigma_{rr}(r,\theta=0)$ associated with in-plane hydrostatic pressure for Examples (ii) and (iii) respectively, whereas (c)-(d) and (e)-(f) illustrate $\sigma_{rr}(r,\theta=0)$ and $\sigma_{r\theta}(r,\theta=\pi/4)$ associated with in-plane shear, for Examples (ii) and (iii) respectively. Far field loading is normalised to unity. 
}
\label{fig2}
\end{figure}

{In Figure \ref{fig2} we plot radial and shear stress distributions as functions of $r$ associated with Example (ii) (left of the figure) and (iii) (right of the figure) in Table \ref{-6}. For all cases we fix the fibre and exterior (matrix) properties and plot the isolated fibre (no coating) case (blue long dash), the weak neutral inclusion case (black solid)  and the strong neutral inclusion case (red short dash). Weak NI properties are deduced from the Christensen and Lo shear conditions (see Appendix \ref{CLapp}) coupled with the standard isotropic bulk modulus condition \eqref{3=5} with $\beta=1$ (isotropic coating). We thus deduce that for example (ii) isotropic coating properties are $K^{\textnormal{weak}}=2.32912, \mu^{\textnormal{weak}}=0.715023$ whereas for  example (iii) isotropic coating properties are $K^{\textnormal{weak}}=2.47029, \mu^{\textnormal{weak}}=0.277903$.}

{Figure \ref{fig2}(a) and (b) illustrate $\sigma_{rr}(r)$ for the \textit{hydrostatic problem} (independent of $\theta$) for Examples (ii) and (iii) respectively. The problem is scaled such that $\sigma_{rr}\rightarrow 1$ as $r\rightarrow\infty$. Figure \ref{fig2}(c)-(e) illustrate the \textit{shear problem ($\sigma_{xx}+\sigma_{yy}=0$ as $r\rightarrow\infty$)}. (c) and (e) correspond to Example (ii) whereas (d) and (f) correspond to Example (iii). $\sigma_{rr}$ is evaluated at $\theta=0$ and $\sigma_{r\theta}$ is evaluated at $\theta=\pi/4$.}

{One should note that strong NIs ensure that the field is unperturbed in $r\geq1$ for the shear problem. The hydrostatic problem is unperturbed for $r\geq1$ for both strong \textit{and} weak NIs as expected. The effect is more noticeable in Example (ii) (void) than for Example (iii) (rigid core). In the latter case the weak NI can be seen as almost as effective as the strong NI. In the former however the weak NI is ineffective in shear. In Example (iii) we also note that the strong NI has a strong influence on the stress distribution in the core.}

\section{Conclusions}  \label{sec7}
The connection between low frequency transparency of elastic waves and neutral inclusions has been made for the first time.  Intuitively, both effects are related  to static or quasi-static cloaking, although as we have seen, the relationships require careful definitions of both neutral inclusions and low frequency transparency.  Two distinct types of neutral inclusion have been identified, weak and strong, with the former equivalent to low frequency transparency and the latter with static cloaking.   The main results of the paper are summarised in Theorem 1 which shows that weak NI/low frequency transparency is easier to achieve than strong NI/static cloaking.   The former can be obtained with an isotropic shell surrounding the core, while the latter requires anisotropy in the shell/cloak.  For a given core and matrix, and relative shell thickness, the determination of the shell properties for either the weak or strong NI effect is implicit through effective medium conditions.  The existence of solutions is not guaranteed, but depends upon the parameters in a non-trivial manner. 


The problem has been made tractable by considering the $n=0, \ 1, \ 2$ sub-problems, with $n=1$ trivially related to density.  The concepts of  low frequency wave transparency and  neutral inclusion are identical for the $n=0$ problem, for which there is no distinction between weak and strong NI effects.   Thus, if the  exterior bulk modulus matches the effective bulk modulus of the core-shell composite cylinder then the latter acts as a neutral inclusion and is transparent in the long wavelength regime.  Distinguishing between weak and strong NI effects are necessary for the $n=2$ problem.  For the weak NI effect the shell properties must be such that the single condition \eqref{53} holds, in which case the effective shear modulus of the shell plus core is given by \eqref{+1}. The strong NI effect requires that Eqs.\  \eqref{53} and \eqref{35} are both satisfied, with matrix effective shear modulus of Eq.\ 
\eqref{36}.

These connections between low frequency transparency, static cloaking and neutral inclusions provide the material designer with options for achieving elastic cloaking in the quasi-static limit.  Extension of the 
results to spherical geometries is the natural next step and will be the subject of a future report.

\appendix
\section{Elastodynamic scattering solution} \label{appA}

Based on the representation \eqref{-1}, let
\beq{1=2}
\begin{aligned}
\phi &= \big( B_L  J_n(k_L r) + D_L H_n^{(1)} (k_L r) \big) k_L^{-1} \de^{\i n \theta},
\\
\psi &=  \big( B_T  J_n(k_T r) + D_T H_n^{(1)} (k_T r) \big) k_T^{-1} \de^{\i n \theta},
\end{aligned}
\eeq
 then, dropping the $\de^{\i n \theta}$ terms,
\bse{1=3}
\bal{1=3a}
{\bf u} &=
{\bf U}_n(J_n,r) {\bf b} + {\bf U}_n(H_n^{(1)},r) {\bf d} ,
\\
r {\bf t} &=
{\bf V}_n(J_n,r)
{\bf b}  + {\bf V}_n(H_n^{(1)},r) {\bf d}  \  \ \text{where}
\\
{\bf u} &= \begin{pmatrix}
u_r
\\
u_\theta
\end{pmatrix} , \ \
{\bf t} = \begin{pmatrix}
t_r
\\
t_\theta
\end{pmatrix} ,
\ \
{\bf b} =\begin{pmatrix}
 B_L
\\
B_T
\end{pmatrix}  , \ \
{\bf d} =\begin{pmatrix}
 D_L
\\
D_T
\end{pmatrix}  ,
\\
{\bf U}_n(f,r) &=
\begin{pmatrix}
 f'(k_Lr)  & -\frac{\i n}{k_Tr} f(k_T r)
\\
\frac{\i n}{k_Lr} f(k_L r) & f'(k_Tr)
\end{pmatrix},
\\
{\bf V}_n(f,r) &=
{\mu_e}
\begin{pmatrix}
k_Lr\big[ 2 f'' (k_Lr) + \big(2 - \frac{k_T^2}{k_L^2}\big)f(k_L r)  \big]
 &  - 2\i n \big[  f' (k_Tr)  - \frac 1{k_Tr}f(k_T r)  \big]
\\
2\i n \big[  f' (k_Lr)  - \frac 1{k_Lr}f(k_L r)  \big]  &
-2f'(k_Tr)   + \big( \frac{2n^2}{k_Tr} - k_Tr \big) f(k_T r)
\end{pmatrix}.
\eal
\ese
Following the notation of  \cite{Norris10}, assume
\beq{1=4}
r {\bf t} = -{\bf Z}_1 {\bf u}\ \text{at} \ r=r_1,
\eeq
then the scattered L and T amplitudes ${\bf d} $ of azimuthal order $n$ can be found in terms of the incident ones ${\bf b} $ as
\beq{1=5}
{\bf d} = - \big(  {\bf V}_n(H_n^{(1)},r_1)  +{\bf Z}_1{\bf U}_n(H_n^{(1)},r_1)   \big)^{-1}
\big(  {\bf V}_n(J_n,r_1)  +{\bf Z}_1{\bf U}_n(J_n,r_1)   \big) {\bf b}
.
\eeq
This is the basic equation for solving the scattering.

The impedance ${\bf Z}_1$ is found by first forming the core impedance, which follows from  \cite[eq.\ (8.9)]{Norris10}.  This serves as the initial condition for integrating the impedance from $r=r_0$ to $r_1$.  Direct integration of the dynamic analog of the Riccati equation \eqref{22} is unstable, however, fast and stable methods exist to circumvent this difficulty.  We use the M\"obius scheme based on eqs.\  (17) and (32) of \cite{Norris2013}.

\section{{Christensen and Lo's Weak Neutral Inclusion}}\label{CLapp}

{Christensen and Lo \cite{Christensen79, christensen1986erratum} developed the so-called Generalised Self-Consistent method and thus provided the following expressions for the effective (exterior) shear modulus $\mu_e$ in terms of the core $\mu_0$ and shell $\mu$ properties. We have re-written these here since we specify core and exterior properties and solve for coating properties. We have also corrected the typographical errors that appeared in the original paper:}
{\begin{align}
D\left(\frac{\mu}{\mu_e}\right)^2+B\frac{\mu}{\mu_e}+A &= 0
\end{align}
where
\begin{align}
D &= 3f(1-f)^2 \left(\frac{\mu_0}{\mu}-1\right)\left(\frac{\mu_0}{\mu}+\eta_0\right) \nonumber\\
& \hspace{0.5cm}+ \left(\frac{\mu_0}{\mu}\eta+\left(\frac{\mu_0}{\mu}-1\right)f+1\right)\left(\frac{\mu_0}{\mu}+\eta_0+\left(\left(\frac{\mu_0}{\mu}\right)\eta-\eta_0\right)f^3\right), \\
B &= -6f(1-f)^2 \left(\frac{\mu_0}{\mu}-1\right)\left(\frac{\mu_0}{\mu}+\eta_0\right) \nonumber\\
& \hspace{0.5cm}+ \left(\frac{\mu_0}{\mu}\eta+\left(\frac{\mu_0}{\mu}-1\right)f+1\right)\left((\eta-1)\left(\frac{\mu_0}{\mu}+\eta_0\right)-2\left(\left(\frac{\mu_0}{\mu}\right)\eta-\eta_0\right)f^3\right), \nonumber\\
& \hspace{0.5cm} + (\eta+1)f\left(\frac{\mu_0}{\mu}-1\right)\left(\frac{\mu_0}{\mu}+\eta_0+\left(\frac{\mu_0}{\mu}-\eta_0\right)f^3\right) \\
A &= 3f(1-f)^2 \left(\frac{\mu_0}{\mu}-1\right)\left(\frac{\mu_0}{\mu}+\eta_0\right) \nonumber\\
& \hspace{0.5cm}+ \left(\frac{\mu_0}{\mu}\eta+\eta_0\eta+\left(\frac{\mu_0}{\mu}-\eta_0\right)f^3\right)\left(\left(\frac{\mu_0}{\mu}-1\right)\eta-\left(\frac{\mu_0}{\mu}\eta+1\right)\right),  
\end{align}}
{with  $f = \frac{r_0^2}{r_1^2}$, $\eta=1+2\dfrac{\mu}{K}$ and $\eta_0=1+2\dfrac{\mu_0}{K_0}$.}

%
%
\subsection*{Acknowledgments} {Thanks to Dr.\ Xiaoming Zhou and Prof.\ Gengkai Hu for useful discussions.} %
{ANN gratefully acknowledge support from the National Science Foundation EFRI Award No. 1641078. WJP thanks the Engineering and Physical Sciences Research Council for supporting this work (via grants EP/L018039/1 AND EXTENSION).}



\end{document}